\documentclass[aps, pre, reprint, superscriptaddress, nofootinbib]{revtex4-2}

\usepackage{graphicx}  
\usepackage{amsmath}   
\usepackage{amssymb}   
\usepackage{hyperref}
\hypersetup{colorlinks=true, linkcolor=blue, citecolor=blue, urlcolor=blue}
\usepackage{dcolumn}   
\usepackage{bm}        
\usepackage{ragged2e}
\usepackage[percent]{overpic}
\usepackage{booktabs}
\usepackage{siunitx}

\usepackage[english]{babel}
\usepackage{float} 
\usepackage{booktabs}
\usepackage{array}
\usepackage{siunitx}
\usepackage{placeins}

\begin{document}


\title{Mass-induced Mpemba effect in a polymer--bead system}

\author{Hosung Kwak}
\affiliation{Department of Physics, KAIST, Daejeon 34141, Republic of Korea}

\author{Yongjoo Baek}
\email{y.baek@snu.ac.kr}
\affiliation{Department of Physics and Astronomy \textup{\&} Center for Theoretical Physics, Seoul National University, Seoul 08826, Republic of Korea}

\author{Hawoong Jeong}
\email{hjeong@kaist.edu}
\affiliation{Department of Physics, KAIST, Daejeon 34141, Republic of Korea}
\affiliation{Center for Complex Systems, KAIST, Daejeon 34141, Republic of Korea}

\begin{abstract}
We propose a physically motivated model in which the Mpemba effect is induced by the system's inertia. The model describes a polymer undergoing a denaturation transition whose force–extension curve contains a weak-force plateau that slows relaxation.
In the presence of inertia, a farther initial state can reach and cross the plateau sooner, producing the Mpemba effect. Increasing the bead mass broadens the range of initial conditions over which this mechanism operates. A similar mechanism also generates the inverse Mpemba effect.
\end{abstract}

\maketitle 

\section{Introduction}

The Mpemba effect in its original form refers to the observation that, for some range of temperature, initially hotter water freezes sooner than initially colder water after a thermal quench, contradicting the intuitive expectation based on Newton's law of cooling~\cite{mpemba1969cool}. It is now regarded as a paradigmatic example of anomalous relaxation, and the term ``Mpemba effect'' has been extended to situations in which a state initially farther from equilibrium relaxes faster than one initially closer to
it~\cite{TEZA20261}.

The past decade has seen reports of this generalized Mpemba effect in spin systems~\cite{PhysRevLett.124.060602,vadakkayil2021should,PhysRevE.110.L012103,doi:10.1073/pnas.1819803116}, colloidal particles~\cite{kumar2020exponentially,doi:10.1073/pnas.2118484119,bechhoefer2021fresh}, clathrate hydrates~\cite{ahn2016experimental}, active matter~\cite{PhysRevLett.129.138002,10.1063/5.0246857,PhysRevE.111.054136,j4wb-1d9d}, granular fluids~\cite{PhysRevLett.119.148001,PhysRevE.99.060901,10.1063/5.0050804,PhysRevE.102.012906,Biswas_2021,biswas2022mpemba,megias2022mpemba,PhysRevE.108.024902},  inertial suspensions~\cite{PhysRevE.103.032901}, and even quantum systems~\cite{PhysRevLett.131.080402,Ares2023,PhysRevLett.133.140404,zhang2025observation,PhysRevLett.134.220402,rbt4-psfd,yzjd-pk8h,hallam2025tunablequantummpembaeffect,shah2026tunablempembaeffectprethermal}. The diversity of these examples demonstrates that anomalous relaxation is a generic feature of nonequilibrium dynamics rather than a curious exception, providing a feasible basis for implementing shortcuts to a target state by choosing appropriate initial conditions or protocols~\cite{PhysRevLett.124.060602,PhysRevLett.132.117102}.

Such ubiquity and utility of the Mpemba effect naturally raise the question of what makes it occur. Notably, a theoretical study based on a Markov jump process addressed this issue using spectral decompositions of the transition rate matrix~\cite{doi:10.1073/pnas.1701264114,PhysRevX.9.021060}. This approach clarifies when the greater magnitude of a parameter change results in the smaller initial amplitude of the slow relaxation mode, providing sharp criteria for the Mpemba effect in terms of detailed transition rates between states. However, it is not always straightforward to translate such microscopic criteria into mesoscopic or macroscopic design rules~\cite{PhysRevLett.132.117102}. To control anomalous relaxation in a practically feasible manner, we should be able to engineer the Mpemba effect by tuning a small number of macroscopic knobs rather than by fine-tuning microscopic transition rates.

\begin{figure}[b]
  \centering
  \includegraphics[width=\columnwidth]{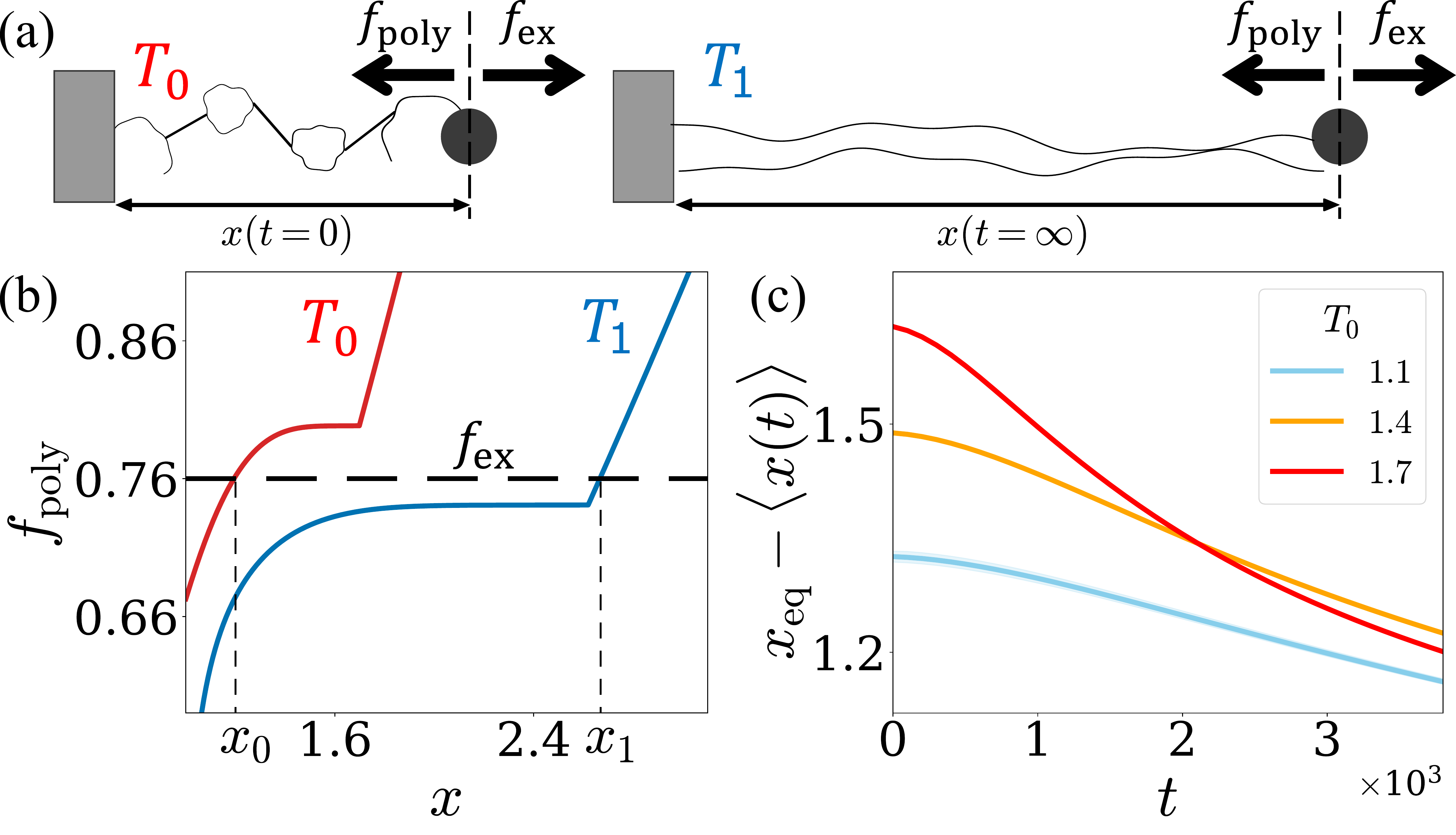}
  \caption{\textbf{(a)} Schematic of the polymer--bead model and the thermal quenching protocol. One end of the polymer is fixed, while the other is attached to a bead of mass $m$ pulled by a constant external force $f_{\mathrm{ex}}$. The bath temperature is instantaneously reduced from $T_0$ to $T_1$ at $t=0$, and the DNA extension $x$ subsequently relaxes from a value $x(0)$ sampled from the initial equilibrium distribution towards the mean value $x_\mathrm{eq}=\langle x(\infty)\rangle$ of the final equilibrium distribution. \textbf{(b)} Temperature-dependent force--extension relations $f_{\mathrm{poly}}(x;T)$ at the initial temperature $T_0$ (red) and at the final temperature $T_1$ (blue), shown together with the constant applied force $f_\mathrm{ex}$ (dashed line). The intersections with $f_\mathrm{ex}$ define the mechanical equilibrium at each temperature: $x_0$ at $T_0$ and $x_1$ at $T_1$. \textbf{(c)} Relaxation behaviors for different values of $T_0$ (with $T_1 = 0.4$ fixed) indicated by the time evolutions of $x_{\mathrm{eq}}-\langle x(t)\rangle$. The relaxation curve for $T_0 = 1.7$ overtakes that of $T_0 = 1.4$, signaling the Mpemba effect.}
  \label{fig:big}
\end{figure}

The central finding of this study is that inertia can serve as a directly tunable control parameter for anomalous relaxation. To demonstrate this, we investigate the Mpemba effect associated with DNA denaturation via overstretching. The key ingredients of our model are (i) a constant stretching force applied on the DNA, (ii) a broad plateau in the force--extension relation of the DNA signaling a stretching-induced denaturation, and (iii) an underdamped dynamics of the DNA extension due to a point mass (bead) attached to its tip. See Fig.~\ref{fig:big}(a, b) for a schematic of this setting, where $f_\mathrm{ex}$, $f_\mathrm{poly}$, and $x$ denote the stretching force, restoring force, and polymer extension, respectively, all of which are made dimensionless by suitable rescaling that will be discussed later.

In this setting, a temperature quench from $T_0$ to $T_1$ (with $T_1 < T_0$) produces a stretching of the DNA, {\em i.e.}, the mechanical equilibrium point (where $f_\mathrm{ex} = f_\mathrm{poly}$ holds) increases from $x_0$ to $x_1$. When the relaxation trajectory crosses the denaturation plateau, and if $f_\mathrm{ex}$ is barely above the plateau of $f_\mathrm{poly}$ as shown in Fig.~\ref{fig:big}(b), the net force $f_\mathrm{ex} - f_\mathrm{poly}$ driving the relaxation becomes very weak. Then, the DNA crosses the plateau regime more quickly when it has already accumulated sufficiently larger velocity by starting farther away from the plateau ({\em i.e.}, near a smaller $x_0$), which corresponds to having a larger initial temperature $T_0$. This simple mechanical picture hints at how we can engineer the Mpemba effect by tuning the mass attached to the DNA, as illustrated by simulations of relaxation towards $x_\mathrm{eq} \equiv \langle x(t=\infty) \rangle$ shown in Fig.~\ref{fig:big}(c).

We note that previous studies have also discussed the Mpemba effect in underdamped systems, such as granular gases~\cite{PhysRevLett.119.148001, PhysRevE.99.060901, 10.1063/5.0050804, PhysRevE.102.012906, Biswas_2021, biswas2022mpemba, megias2022mpemba, PhysRevE.108.024902}, inertial suspensions~\cite{ PhysRevE.103.032901}, and active matter~\cite{j4wb-1d9d}. These studies, however, use inertia only to define effective temperature from kinetic energy and do not explore how inertia can serve as a macroscopic knob controlling the Mpemba effect. More recently, the Mpemba effect was investigated directly for an underdamped Brownian particle in a double-well potential~\cite{Shapira2026}. In the system, inertia tends to suppress the Mpemba effect, so the study focused on whether the effect persists away from the overdamped limit. In contrast, our study demonstrates that mass can {\em induce} the Mpemba effect, with higher mass increasing the parameter range over which the effect is observed.



We organize the rest of this paper as follows. In Sec.~\ref{sec:methods}, we first describe the DNA model, the coarse-grained bead dynamics, and the thermal and mechanical quenching protocols used in the simulations. In Sec.~\ref{sec:results}, we then present numerical results demonstrating the Mpemba and inverse Mpemba effects, exploring how inertia controls the strength and extent of the Mpemba region. In Sec.~\ref{sec:Discussion}, we discuss whether the mechanisms of the tunable Mpemba effect identified by our simulations can be implemented in experimental settings.

\section{Model and Methods}
\label{sec:methods}

\subsection{Stretched DNA model with a force plateau}

Our system of interest is a three-dimensional double-stranded DNA stretched along the $x$-direction by fixing one end of a strand to the wall and the other end to a bead pulled at constant force, as illustrated in Fig.~\ref{fig:big}(a). We describe the system using the Hanke--Ochoa--Metzler model~\cite{PhysRevLett.100.018106}, which generalizes the seminal Poland--Scheraga model describing the thermal denaturation of DNA~\cite{10.1063/1.1727785} to the case of stretching-assisted denaturation. More specifically, the model assumes that the DNA consists of bound double-stranded segments alternating with denatured, single-stranded segments called bubbles. The bound segments are like rigid rods, whereas the bubbles are flexible and can be stretched more easily. As the stretching force on the DNA gets stronger, the model predicts that the bubbles account for an increasing portion of the DNA, resulting in a continuous phase transition from the bound phase to the denatured phase when the bound segments take up only a vanishing fraction of the base pairs.

Notably, the transition accompanies a plateau-like regime in the force--extension relation: at a given temperature $T$, the magnitude of the DNA's restoring force $f_\mathrm{poly}(x;T)$ exhibits a broad regime in which the force varies only weakly with extension $x$ until it abruptly changes the slope at the phase boundary, as illustrated in Fig.~\ref{fig:big}(b). For a sketch of how one can derive $f_\mathrm{poly}(x;T)$ from the model, we refer the reader to Appendix~\ref{app:HOM}.

We note that the derivation of $f_\mathrm{poly}(x;T)$ assumes the DNA to be at thermal equilibrium for the given values of $x$ and $T$. Throughout this study, we assume that $f_\mathrm{poly}(x;T)$ accurately describes the DNA's restoring force even as $x$ changes in time. The validity of this assumption will be discussed in Sec.~\ref{sec:Discussion}.


\subsection{Underdamped bead dynamics}




The assumptions stated above allow us to describe the dynamics of the system solely in terms of the DNA extension $x$. To incorporate inertia into the dynamics, we attach a bead of mass $m$ to the free end of the polymer, so that $x$ also indicates the dimensionless position of the bead rescaled by the number of base pairs in the DNA, $N$. We further assume that the bead is subject to an externally applied stretching force $f_{\mathrm{ex}}$, which can be realized experimentally by optical tweezers. Then, the underdamped relaxation dynamics of the bead is described by the Langevin equation
\begin{align} \label{eq:langevin}
    mN\ddot{x} = f_\mathrm{ex}-f_\mathrm{poly}(x;T)-N\dot{x}+\sqrt{2T}\xi,
\end{align}
where $\xi$ is a Gaussian white noise with $\langle \xi(t) \rangle = 0$ and $\langle \xi(t)\xi(t') \rangle = \delta(t-t')$. All quantities appearing in this equation have been made dimensionless by choosing the base-pair spacing $a$ as the length unit, the base-pair binding energy $\varepsilon$ as the energy unit, and the drag coefficient $\gamma$ to be unity. This is also equivalent to choosing $\varepsilon/a$ as the unit of force, $\varepsilon/k_{\mathrm B}$ as the unit of temperature, $\gamma a^2/\varepsilon$ as the unit of time, and $\gamma^2 a^2/\varepsilon$ as the unit of mass.

To numerically integrate Eq.~\eqref{eq:langevin}, we use the ``BAOAB'' method, which is one of the standard schemes for simulating an underdamped Langevin system~\cite{10.1063/1.4802990}, with the discretized time step given by $dt = 1$. During the simulation, we evaluate the polymer force $f_{\mathrm{poly}}(x;T)$ by interpolating the pre-computed values on a grid.

Through all simulations, we fix $N = 1000$. The values of the other parameters will be specified in the captions of the relevant figures.

\subsection{Equilibration and quenching protocol}
\label{sec:relax}

The simulation consists of three stages: (i) preparation of the initial equilibrium state, (ii) relaxation after a temperature/force quench, and (iii) observation of the final equilibrium state.

To prepare the initial equilibrium state at given initial temperature and stretching force, we put the bead at the mechanical equilibrium point $x = x_0$ and let it evolve according to Eq.~\eqref{eq:langevin} for $10^5$ time steps. We record the phase-space trajectory of the bead during the last $5 \times 10^4$ time steps of the equilibration stage as a representation of the initial equilibrium ensemble.

The relaxation stage starts by changing either the bath temperature from $T_0$ to $T_1$ (while the stretching force stays at $f_\mathrm{ex}$) or the stretching force from $f_0$ to $f_1$ (while the bath temperature stays at $T$) at $t = 0$. To implement this quenching effect, the bead position and momentum are sampled from the initial equilibrium ensemble described above. Then, the bead evolves according to Eq.~\eqref{eq:langevin} for $10^6$ time steps. We repeat this relaxation stage to obtain an ensemble of $10^5$ trajectories and compute the ensemble-averaged position $\langle x(t) \rangle$ and kinetic energy $\langle K(t)\rangle = \langle \frac{1}{2}m N^2\dot{x}(t)^2 \rangle$ at each time step. To measure how different regimes of the $f_\mathrm{poly}$ landscape contribute to the relaxation time, we operationally define the interval of $x$ with $|df_\mathrm{poly}/dx| < 0.05$ as the \textit{plateau regime}, and separately measure the lengths of time $\langle x(t) \rangle$ stays in the pre-plateau regime (\textit{i.e.}, the initial regime before the bead enters the plateau) and in the plateau regime, which we denote by $t_\mathrm{pre}$ and $t_\mathrm{plat}$, respectively.

Finally, to determine when the system achieves thermal equilibrium, we separately simulated Eq.~\eqref{eq:langevin} from the mechanical equilibrium point $x = x_1$ for $10^6$ time steps and calculated the mean position $x_\mathrm{eq}$, the position variance $\sigma_x^2$, and the kinetic energy variance $\sigma_K^2$ during the last $10^5$ time steps. Given these values, we define the relaxation time $t_{\mathrm{relax}}$ as the time at which both $\langle x(t)\rangle$ and $\langle K(t)\rangle$ enter the tolerance windows around their equilibrium values given by
\begin{align}
  \bigl|\langle x(t)\rangle - x_{\mathrm{eq}}\bigr| < 3\sigma_x
  ~\text{and}~
  \bigl|\langle K(t)\rangle - \tfrac{1}{2}k_{\mathrm B}T_1\bigr| < 3\sigma_K.
\end{align}
We note that similar criteria based on the state of the system or the kinetic energy have been used in previous studies~\cite{PhysRevLett.119.148001,PhysRevE.103.032901,ZhangPRE2022}.

To obtain a robust estimate in the presence of thermal noise, we allow brief excursions outside the windows set by the inequalities and define $t_{\mathrm{relax}}$ as the earliest time after which any continuous excursion outside the windows lasts less than $2\times 10^4$ time steps, which is comparable to the relaxation time scale expected from Eq.~\eqref{eq:langevin} and the force--extension relation shown in Fig.~\ref{fig:big}(b). This avoids misclassifying an already equilibrated trajectory simply because of spontaneous fluctuations.

\section{Results}
\label{sec:results}

\subsection{Mpemba effect and the role of inertia}
\label{sec:Mpemba}

\begin{figure}
\centering
  \includegraphics[width=\columnwidth]{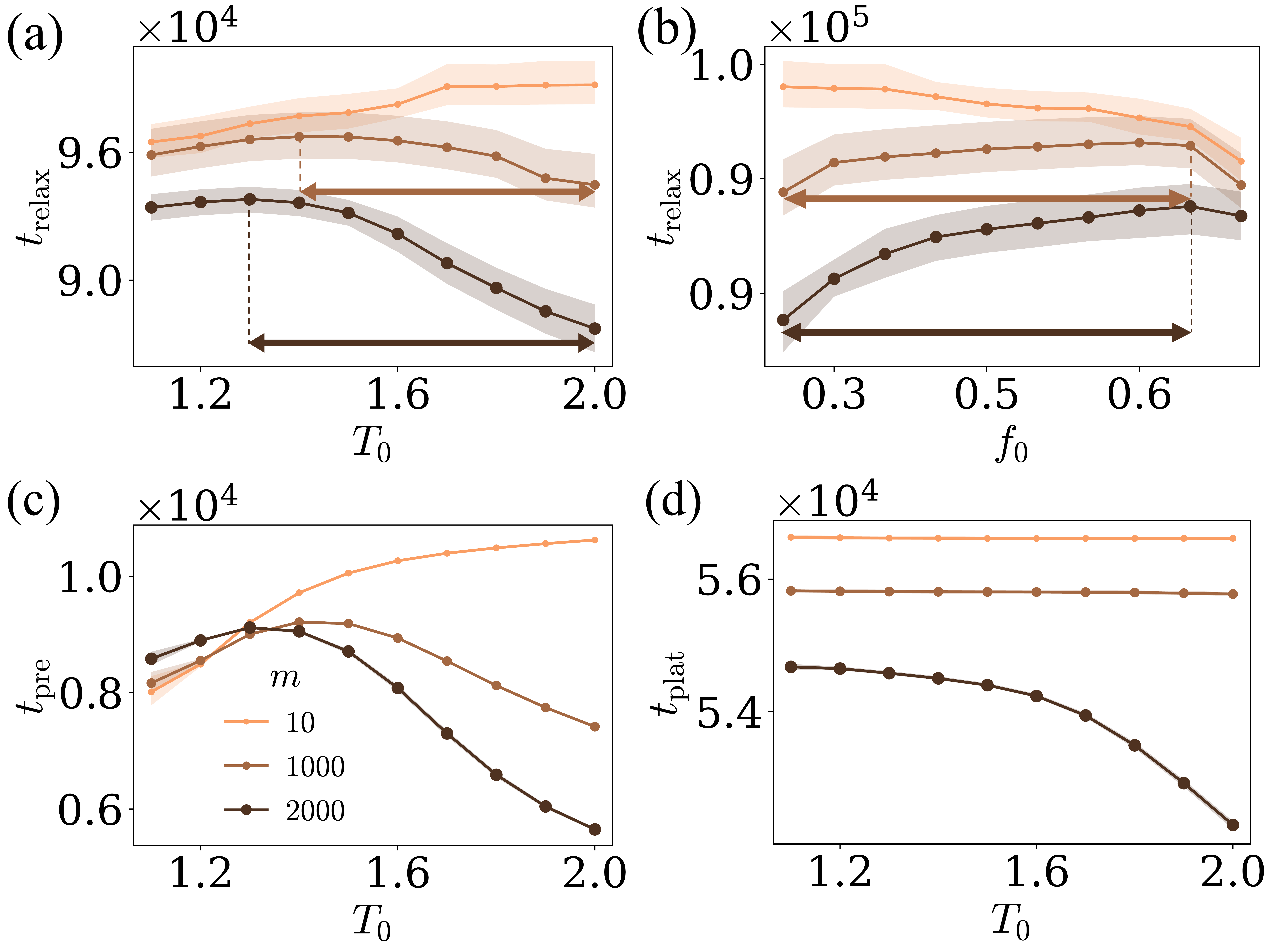}
    \caption{
    Simulation results demonstrating the Mpemba effect induced by a point mass attached to an overstretched DNA.
    Relaxation times $t_\mathrm{relax}$
    \textbf{(a)} for a temperature quench from $T_0$ to $T_1=0.4$ with fixed $f_\mathrm{ex} = 0.76$ and
    \textbf{(b)} for a force quench from $f_0$ to $f_1 =0 .76$ with fixed $T = 0.4$ are shown for different masses $m$. The ranges of $T_0$ and $f_0$ exhibiting the Mpemba effect are highlighted by horizontal arrows. We also show contributions of
    \textbf{(c)} the pre-plateau regime and
    \textbf{(d)} the plateau regime to $t_\mathrm{relax}$ after the temperature quench shown in (a). 
    Lines are to guide the eye, and shaded bands represent the $95\%$ bootstrap confidence intervals.
    }
  \label{fig:mpemba_mass}
\end{figure}

Now, we discuss the results of our simulations demonstrating the Mpemba effect induced by a point mass attached to an overstretched DNA. In Fig.~\ref{fig:mpemba_mass}(a), we show the relaxation time $t_\mathrm{relax}$ after a sudden temperature drop from $T_0$ to $T_1 = 0.4$ while the stretching force remains fixed at $f_\mathrm{ex} = 0.76$. When the mass is small ($m = 10$), $t_\mathrm{relax}$ increases monotonically over the observed range of $T_0$, not showing any signal of the Mpemba effect. It should be noted that we do observe a decrease of $t_\mathrm{relax}$ with $T_0$ in the low-temperature regime not shown in Fig.~\ref{fig:mpemba_mass}(a), but this is due to the nonmonotonic $T_0$-dependence of $x_0$, not the effects of the mass; see Appendix~\ref{app:geo} for more detail. In contrast, when $m = 1000$ or larger, there appears a range of $T_0$ (indicated by a horizontal arrow, which we call the ``Mpemba region'') where increasing $T_0$ reduces $t_\mathrm{relax}$. The Mpemba region broadens further for $m = 2000$, verifying that the effect is tunable by varying the inertia.

As Fig.~\ref{fig:mpemba_mass}(b) shows, we observe similar phenomena when the stretching force $f_\mathrm{ex}$ suddenly increases from $f_0$ to $f_1 = 0.76$ for a fixed temperature $T = 0.4$. Since such force change moves the mechanical equilibrium point from $x_0$ to $x_1$ across the $f_\mathrm{poly}$ plateau in a manner fully analogous to the temperature quench, we can naturally expect that this case would also exhibit a similar decrease of $t_\mathrm{relax}$ as the force change $f_1 - f_0$ increases. Accepting the extended definition of the Mpemba effect as the faster relaxation achieved by a greater change of a control parameter, we also regard the results of this mechanical quenching protocol as a demonstration of the Mpemba effect, as done by some previous studies in different settings~\cite{Degunther2022, PhysRevE.103.032901, Ares2023}.

In Figs.~\ref{fig:mpemba_mass}(c) and (d), we show how the pre-plateau and the plateau regimes of the $f_\mathrm{poly}$ landscape contribute to $t_\mathrm{relax}$ after the temperature quench shown in Fig.~\ref{fig:mpemba_mass}(a), respectively. When the mass is small ($m = 10$), $t_\mathrm{pre}$ increases monotonically with $T_0$, while $t_\mathrm{plat}$ varies little with $T_0$. This indicates that, as damping overwhelms inertia, the bead does not accelerate much in the pre-plateau regime and enters the plateau regime with its velocity already very close to the terminal velocity, which is independent of $T_0$. In this case, one cannot reduce $t_\mathrm{relax}$ by placing the bead farther away from the plateau; hence the absence of the Mpemba effect.

In contrast, when $m = 1000$, we observe that $t_\mathrm{pre}$ does not increase monotonically with $T_0$. The results show that, if $T_0 \gtrsim 1.3$, putting the bead farther away from the plateau by increasing $T_0$ can make the bead reach the plateau more quickly. This stems from the curvature of $f_\mathrm{poly}$ in the pre-plateau regime. As illustrated in Fig.~\ref{fig:big}(b), $f_\mathrm{poly}$ is a concave function of $x$ in the pre-plateau regime. This implies that, if we neglect the damping, the initial relaxation behavior of the bead is similar to the dynamics of an anharmonic oscillator with positive quartic terms in the potential energy. For such oscillators, the period of oscillation decreases with amplitude (in contrast to harmonic oscillators whose period is independent of amplitude), which explains the decrease of $t_\mathrm{pre}$ with $T_0$ for $T_0 \gtrsim 1.3$. On the other hand, the same explanation is not applicable when $m = 10$ (small mass) or $T_0 \lesssim 1.3$ (weak anharmonic force) since the damping effects are more dominant in these cases. As for the plateau regime, $t_\mathrm{plat}$ for $m = 1000$ again shows little dependence on $T_0$, indicating that the bead moves at terminal velocity throughout the plateau regime. Thus, the Mpemba effect observed for $m = 1000$ is mainly due to the faster crossing of the pre-plateau regime.

When $m = 2000$, in addition to a similar nonmonotonic behavior of $t_\mathrm{pre}$, we observe that $t_\mathrm{plat}$ decreases significantly as $T_0$ increases. In this case, the bead enters the plateau regime at a velocity larger than the terminal velocity. Even though the difference between these two velocities will soon converge to zero, if the inertia is large enough to delay the convergence, having a larger entrance velocity can significantly lower the time it takes to cross the plateau. For $m = 2000$, it turns out that both $t_\mathrm{pre}$ and $t_\mathrm{plat}$ make comparable contributions to the Mpemba effect.

\subsection{Inverse Mpemba effect}

\begin{figure}[t]
  \centering
  \includegraphics[width=\columnwidth]{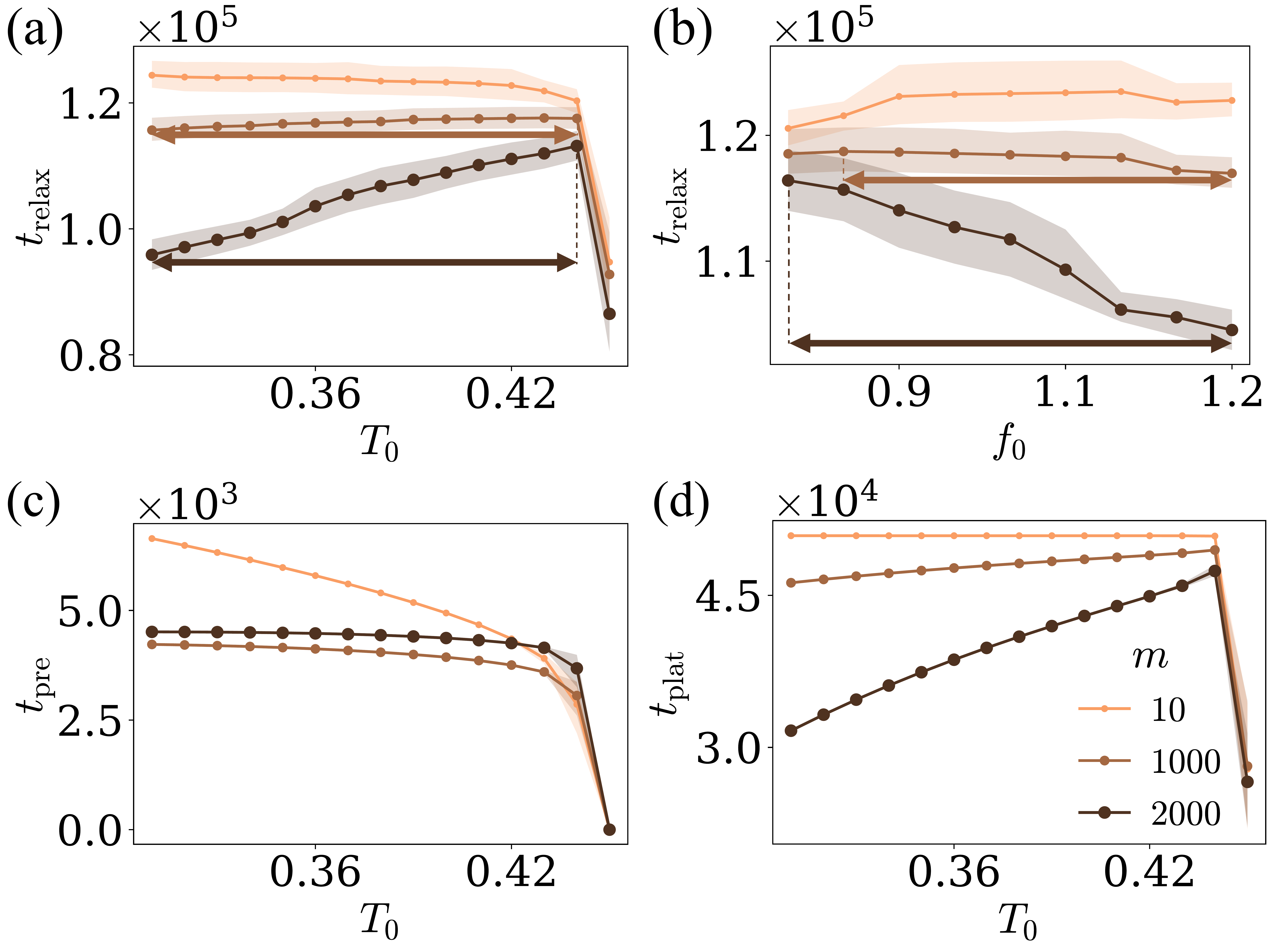}
    \caption{
    Simulation results demonstrating the inverse Mpemba effect induced by a point mass attached to an overstretched DNA.
    Relaxation times $t_\mathrm{relax}$
    \textbf{(a)} for a temperature quench from $T_0$ to $T_1=0.5$ with fixed $f_\mathrm{ex} = 0.76$ and
    \textbf{(b)} for a force quench from $f_0$ to $f_1 =0 .76$ with $T = 0.5$ are shown for different masses $m$. The ranges of $T_0$ and $f_0$ exhibiting the inverse Mpemba effect are highlighted by horizontal arrows. We also show contributions of
    \textbf{(c)} the pre-plateau regime and
    \textbf{(d)} the plateau regime to $t_\mathrm{relax}$ after the temperature quench shown in (a). 
    Lines are to guide the eye, and shaded bands represent the $95\%$ bootstrap confidence intervals.
    }
  \label{fig:Inverse_Mpemba}
\end{figure}

We also observe that the plateau-crossing mechanism introduced above can produce an inverse Mpemba effect, \textit{i.e.}, the system equilibrates more quickly when it heats up starting from a lower initial temperature. In Fig.~\ref{fig:Inverse_Mpemba}(a), we show the relaxation time $t_\mathrm{relax}$ after a sudden temperature rise from $T_0$ to $T_1 = 0.5$ with the stretching force fixed at $f_\mathrm{ex} = 0.76$. For $m = 10$, $t_\mathrm{relax}$ increases monotonically as $T_0$ is lowered away from $T_1$; no inverse Mpemba effect is observed in this nearly overdamped regime. In contrast, when $m = 1000$ or larger, there appears a range of $T_0$ (indicated by a horizontal arrow and referred to as the ``inverse Mpemba region’’) in which lowering $T_0$ farther away from $T_1$ reduces $t_\mathrm{relax}$. The same effect becomes even more pronounced for $m=2000$, confirming that it can be tuned by inertia.

In Fig.~\ref{fig:Inverse_Mpemba}(b), we plot $t_\mathrm{relax}$ as the stretching force $f_\mathrm{ex}$ suddenly decreases from $f_0$ to $f_1 = 0.76$ for a fixed temperature $T = 0.5$. The results demonstrate a mechanical version of the inverse Mpemba effect: the system equilibrates more quickly when it shrinks starting from a stronger initial stretching force. Again, the effect is more pronounced when the bead mass becomes larger.

In Figs.~\ref{fig:Inverse_Mpemba}(c) and (d), we show how the pre-plateau and plateau regimes of the $f_\mathrm{poly}$ landscape contribute to $t_\mathrm{relax}$ after the temperature quench. For the small mass ($m=10$), lowering $T_0$ farther away from $T_1$ increases $t_\mathrm{pre}$, while $t_\mathrm{plat}$ remains almost unchanged (except for the narrow region around $T_0\simeq 0.45$ where the initial equilibrium position already lies within the plateau regime). These show that the bead is effectively overdamped, failing to gain any substantial inertial speedup through the pre-plateau regime and entering the plateau regime already at terminal velocity. This explains why the inverse Mpemba effect is absent for $m = 10$.

In contrast, when $m = 1000$ or larger, $t_\mathrm{pre}$ becomes nearly independent of $T_0$ for $T_0\lesssim 0.44$. This independence stems from the approximate linearity of $f_\mathrm{poly}$ in the corresponding pre-plateau regime. If damping is small enough, the bead dynamics in this region is close to that of a harmonic oscillator, whose period is independent of amplitude (isochronism). This explains why $t_\mathrm{pre}$ remains almost constant in this range of $T_0$. Meanwhile, $t_\mathrm{plat}$ exhibits a marked decrease as $T_0$ is lowered. This is because a lower $T_0$ makes the bead enter the plateau regime with a higher velocity. Let us recall that, for the Mpemba effect, both $t_\mathrm{pre}$ and $t_\mathrm{plat}$ make significant contributions. In contrast, for the inverse Mpemba effect, the reduction of $t_\mathrm{plat}$ plays a dominant role.

\section{Discussion}
\label{sec:Discussion}

\newcommand{\twoline}[2]{%
  \begin{tabular}[c]{@{}l@{}}#1\\#2\end{tabular}%
}

\begin{table}
\centering
\caption{(Top) Realistic values of parameters used in the model. (Middle) Values of key observables compatible with the chosen units. (Bottom) Reference values for comparison with realistic experimental settings.}
\label{tab:physical_scales}

\begin{tabular}{lll}
\toprule
Quantity & Notation & Value \\ 
\midrule

\twoline{Base-pair spacing}{(length unit)}
& $a$
& $0.34~\mathrm{nm}$ \\[4pt]

\twoline{Binding energy}{(energy unit)}
& $\varepsilon$
& $10^{-20}\,\mathrm{J}$ \\[4pt]

\twoline{Friction coefficient}{($1\,\mu\mathrm{m}$-radius bead in water)}
& $\gamma$
& $2 \times 10^{-8}\,\mathrm{kg/s}$ \\[4pt]

Mass unit
& $m_0=\gamma^2 a^2/\varepsilon$
& $4.6\times10^{-16}\,\mathrm{kg}$ \\[2pt]

Time unit
& $t_0=\gamma a^2/\varepsilon$
& $2.3\times10^{-8}\,\mathrm{s}$ \\[2pt]

Temperature unit
& $\varepsilon/k_\mathrm{B}$
& $724~\mathrm{K}$ \\[2pt]

\midrule

\twoline{Initial-state temperature}{(in Fig.~\ref{fig:Inverse_Mpemba})}
& $T_0$
& $224~\mathrm{K}$--$326~\mathrm{K}$ \\[4pt]

\twoline{Final-state temperature}{(in Fig.~\ref{fig:Inverse_Mpemba})}
& $T_{\mathrm b}$
& $362~\mathrm{K}$ \\[4pt]

\twoline{Initial external force}{(in Fig.~\ref{fig:Inverse_Mpemba})}
& $f_0$
& $31.6~\mathrm{pN}$--$47.4~\mathrm{pN}$ \\[4pt]

Final external force
& $f_1$
& $30~\mathrm{pN}$ \\[2pt]

Net force in the plateau regime
& --
& $3~\mathrm{pN}$ \\[2pt]

Bead mass
& $1000\,m_0$
& $4.6 \times 10^{-13}~\mathrm{kg}$ \\[2pt]

Bead relaxation time
& $t_\mathrm{relax}$
& $\sim 10^{-3}\,\mathrm{s}$ \\[2pt]

\midrule

\twoline{Liquid water temperature}{($1~\mathrm{atm}$)}
& --
& $273~\mathrm{K}$--$373~\mathrm{K}$ \\[4pt]

\twoline{Bead mass}{(platinum bead of radius $1\,\mu\mathrm{m}$)}
& --
& $8.98 \times 10^{-14}\,\mathrm{kg}$ \\[4pt]

\twoline{DNA relaxation time}{(for micrometer length change)}
& --
& $\sim 1~\mathrm{s}$ \\

\bottomrule
\end{tabular}

\end{table}

Let us first relate our mechanical interpretation of the Mpemba effect to the spectral framework of Refs.~\cite{doi:10.1073/pnas.1701264114,PhysRevX.9.021060}. In the present system, the slow relaxation mode is associated with the bead traversing the plateau of the force--extension curve at approximately its terminal velocity. When the bead mass is small, its velocity rapidly approaches the terminal value, leaving little freedom to alter the subsequent plateau-crossing dynamics through the choice of initial condition. Consequently, the slow mode is difficult to suppress. For a sufficiently large mass, by contrast, the bead retains memory of its pre-plateau acceleration and can enter the plateau with a substantial velocity. An appropriate initial preparation can then reduce the amplitude of the slow mode and accelerate relaxation. The bead mass therefore acts as a macroscopic control parameter that determines how effectively the initial amplitude of the slowest relaxation mode can be tuned.

Although increasing the bead mass enhances the Mpemba effect throughout the parameter range examined here, we expect this trend to break down at sufficiently large masses. Once the bead becomes very heavy, the slowest relaxation process should no longer be its passage across the force plateau, but rather the dissipation of its kinetic energy. In this regime, states prepared farther from the final state may also carry more excess kinetic energy and therefore require more time to relax. Further increasing the mass would then strengthen this inertial bottleneck and ultimately suppress the Mpemba effect. We thus anticipate an optimal intermediate range of masses in which inertia is large enough to facilitate control of the plateau-crossing mode, but not so large that kinetic-energy dissipation becomes the dominant slow process.

We next discuss whether the same mechanism could be observed in an experimental DNA--bead system. Table~\ref{tab:physical_scales} translates the parameters used in our calculations, particularly those of Fig.~\ref{fig:Inverse_Mpemba} for the inverse Mpemba effect, into physical scales relevant to a DNA molecule attached to a bead in water. The initial-temperature interval over which the inverse Mpemba effect occurs overlaps with the range in which water remains liquid. For the mechanical-quench protocol, which may be easier to implement as a true abrupt quench compared to the thermal quench, the required values of the initial force $f_0$ lie entirely within the range accessible using optical tweezers. The principal difficulty concerns the bead mass. For a micron-sized bead immersed in water, the masses at which the Mpemba effects become pronounced in our model are several times larger than realistic values. More fundamentally, the mechanism requires underdamped motion, whereas micron-sized objects in water are typically strongly overdamped. An experimental realization may therefore require an unusually dense micron-sized bead, a larger bead made from a conventional material, or a geometry in which the bead is coupled to the DNA while experiencing substantially less viscous damping than it would if fully immersed in water.

A second experimental challenge concerns the assumed separation of time scales between the polymer and the bead. Our model treats the polymer as relaxing much faster than the bead, allowing its restoring force to be represented at every instant by the gradient of an equilibrium free energy. In realistic DNA--bead systems, however, the internal relaxation time of the DNA may be longer than the characteristic relaxation time of the bead, as discussed in Ref.~\cite{DNArelaxation} and listed in Table~\ref{tab:physical_scales}. Equation~\eqref{eq:langevin} should therefore be regarded as an effective approximation in this setting. Nonequilibrium relaxation of the polymer's internal degrees of freedom may introduce memory effects, modify the force exerted on the bead, and alter both the magnitude and the parameter range of the Mpemba effect. These effects would need to be incorporated when designing and interpreting an experimental implementation.

These practical limitations do not, however, affect the central principle demonstrated by our model. It illustrates that coupling a force landscape containing a weak-driving bottleneck to an underdamped coordinate provides a general route to Mpemba-type relaxation. Inertia allows the system to enter the bottleneck with a preparation-dependent velocity and thereby offers a direct means of tuning the amplitude of the associated slow relaxation mode. This principle is not specific to DNA and would be more broadly applicable.


\begin{acknowledgments}
H. K. and H. J. acknowledge the support by the Basic Science Research Program through the National Research Foundation of Korea (RS-2025-00514776); Y. B. was supported by the National Research Foundation of Korea (NRF) grant funded by the Korea government (MSIT) (No. RS-2023-00278985).
\end{acknowledgments}

\appendix

\section{Hanke--Ochoa--Metzler model}
\label{app:HOM}

Here, for completeness, we sketch how one can obtain the force--extension relation from the Hanke--Ochoa--Metzler model of DNA denaturation.

While we will ultimately consider the case where the DNA consists of a fixed number of base pairs, it is convenient to assume that the number can fluctuate at fugacity $z$. Further assuming that a force of magnitude $F$ is stretching the DNA along the $x$-direction and that the surrounding heat bath has temperature $T \equiv 1/(k_\mathrm{B}\beta)$, one can describe the equilibrium statistics of the system using the grand partition function
\begin{align}
\mathcal{Z}(z, F, T) &= S_\mathrm{e} + S_\mathrm{e} \left[\sum_{n=0}^\infty (BS)^n\right] BS_\mathrm{e}.
\end{align}
Here, the second term on the rhs, which sums to $(BS_\mathrm{e}^2)/(1-BS)$, describes the case where the DNA consists of an alternating sequence of two types of domains: bound double-stranded segments (represented by $B$) and denatured, single-stranded segments called bubbles (represented by $S$). We note that $S_\mathrm{e}$ describes a denatured boundary segment, which does not form a loop due to only one strand being anchored to the wall or the bead, while the tip of the other strand is left to roam freely.

The model assumes each bound segment to be like a rigid rod whose orientation is biased only by the stretching force $F$ in the $x$-direction. Summing over the contribution of a bound segment of $k$ base pairs, we obtain the weight of each bound segment
\begin{align}
B(z,F,T) &= \sum_{k=1}^\infty \frac{(\omega z)^k}{4\pi} \int d\Omega\, \exp(y k \cos\theta) \nonumber\\
&= \frac{1}{2y} \ln \frac{1-\omega z \mathrm{e}^{-y}}{1-\omega z \mathrm{e}^{y}}.
\end{align}
Here, $\omega \equiv \exp(\beta \varepsilon)$ describes the weight contribution of each bound base pair (whose binding energy is $\varepsilon > 0$), $y \equiv \beta F a$ accounts for the effects of the stretching force (with $a$ denoting the distance between adjacent base pairs), and $\int d\Omega$ is the integral over the unit sphere with area $4\pi$ (with $\theta \in [0,\pi]$ corresponding to the angle between the bound segment and the $x$-axis).  

Meanwhile, the model assumes each denatured loop to be like a closed, self-avoiding random walk. Summing over the contribution of a loop of $2l$ monomers ($l$ broken base pairs), we obtain the weight of each loop 
\begin{align}
S(z,F,T) = A \,y^{-1+1/(2\nu)}\, \mathrm{Li}_c [sz \exp(\alpha y^{1/\nu})].
\end{align}
Here, $\nu \approx 0.588$ is the exponent describing the diffusion of a self-avoiding random walker in three dimensions, $c = 4\nu - 1/2 \approx 1.85$ accounts for the self-avoiding random walk forming a closed loop under the influence of the stretching force, $\mathrm{Li}_c (u) = \sum_{l=1}^\infty u^l l^{-c}$ is the polylog function, and $A$, $s$, $\alpha$ are all positive dimensionless coefficients characterizing the amplitude of $S$, the number of possible choices in each random walk step, and the length scale of each random walk step, respectively. Throughout this study, we fix $A = 1$, $\alpha = 1$, and $s = 5$.

Finally, one can obtain the weight $S_\mathrm{e}$ of each denatured boundary segment using a similar relation, although $c$ should be replaced with $\zeta = -0.232$ to account for the segment not being a loop, and $A$ should be replaced with another amplitude $A_\mathrm{e}$:
\begin{align}
S_\mathrm{e}(z,F,T) = A_\mathrm{e} \,y^{-1+1/(2\nu)}\, \mathrm{Li}_\zeta [sz \exp(\alpha y^{1/\nu})].
\end{align}
We note that the value of $A_\mathrm{e}$ does not affect the force--extension relation in the thermodynamic limit.

Within the grand-canonical description introduced above, the number of base pairs, which we denote by $N$, has the mean value
\begin{align}
\langle N \rangle = \frac{\partial}{\partial \ln z} \ln \mathcal{Z}(z,f,T).
\label{eq:number}
\end{align}
Here, $f \equiv Fa/\varepsilon$ is the dimensionless force. Using ensemble equivalence, this allows us to calculate $z$ as a function of $N$ by putting $\langle N \rangle = N$. In the thermodynamic limit \(N\to\infty\), the relevant fugacity \(z^*(f,T)\) is identified as the smallest value of \(z\) at which Eq.~\eqref{eq:number} diverges. This divergence can occur in two distinct ways:
\begin{align}
&B(z^*,f,T)\,S(z^*,f,T) = 1, \label{eq:bound} \\
&z^*(f,T) = \exp\!\bigl(-\alpha y^{1/\nu}\bigr)/s.
\label{eq:denatured}
\end{align}
Between these, if Eq.~\eqref{eq:bound} determines $z^*$, the DNA is in the bound phase where the alternating sequence of bound segments and denatured bubbles accounts for its length. By contrast, if Eq.~\eqref{eq:denatured} determines $z^*$, the DNA is in the denatured phase where the denatured boundary segments take up most of its length. Hence, when the governing equation of $z^*$ switches from Eq.~\eqref{eq:bound} to Eq.~\eqref{eq:denatured}, the DNA undergoes a denaturation transition.

Using the Euler equation, we identify $-(N/\beta) \ln z^*$ to be the grand potential of the DNA. Thus, the mean extension $x$ of the DNA, if normalized by its contour length $Na$, satisfies
\begin{align}
x = \frac{1}{Na}\,\frac{\partial}{\partial F}\left(-\frac{N}{\beta}\ln z^*\right) = -\frac{1}{\varepsilon\beta}\frac{\partial}{\partial f} \ln z^*.
\label{eq:x-f}
\end{align}
This expresses $x$ as a function of the dimensionless force $f$ and temperature $T$. Inverting the relation, we obtain the force--extension relation $f = f_\mathrm{poly}(x;T)$.

\section{Mpemba-like effect unrelated to inertia}
\label{app:geo}

\begin{figure}[t]
  \centering
  \includegraphics[width=\columnwidth]{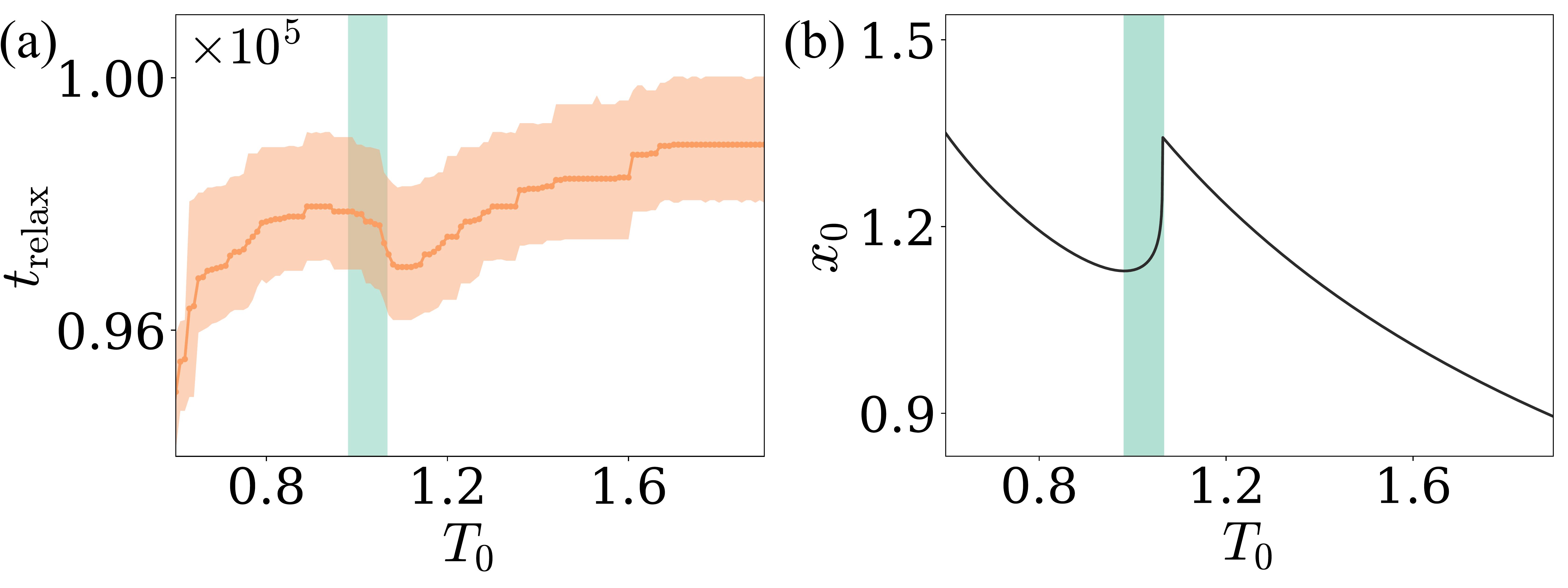}
  \caption{The Mpemba-like effect at small mass ($m=10$).
\textbf{(a)} Relaxation time $t_{\mathrm{relax}}$ versus initial temperature $T_0$, shown over an extended temperature range beyond the main window discussed in Fig.~\ref{fig:mpemba_mass}. A Mpemba-like interval (green shaded) appears even in the nearly overdamped regime. The curve is obtained from five independent random seeds, and the orange shaded band denotes the bootstrap 95\% confidence interval.
\textbf{(b)} Temperature dependence of the equilibrium extension estimated from the force-balance condition $f_{\mathrm{poly}}(x;T_0)=f_{\mathrm{ex}}$, showing a corresponding nonmonotonicity in the shaded window. This implies that increasing $T_0$ can place the initial equilibrium position closer to the final equilibrium extension. The force-balance estimate may differ slightly from the true equilibrium mean due to thermal fluctuations.}
  \label{fig:geometry}
\end{figure}

Our discussions have focused on the Mpemba effect generated by inertia and the plateau in the force--extension relation. For completeness, we comment on an additional mechanism that also produces an Mpemba-like effect but is unrelated to inertia.

As shown in Fig.~\ref{fig:geometry}(a), given appropriate parameters, the relaxation time $t_{\mathrm{relax}}$ can decrease over some intervals of the initial temperature $T_0$ even when the bead mass is small ($m = 10$). However, in this case, the decreasing $t_{\mathrm{relax}}$ is due to the increasing polymer extension $x_0$ over the same interval of $T_0$ as shown in Fig.~\ref{fig:geometry}, which comes from the crossover from a state dominated by double-stranded segments to a state dominated by single-stranded bubbles~\cite{PhysRevLett.100.018106}. Thus, over this $T_0$ interval, a hotter initial state happens to start closer to the final equilibrium position than a colder one. In such cases, the Mpemba-like behaviors should be attributed to the nontrivial equilibrium structures of the DNA, but not the inertia.

\FloatBarrier
\bibliography{apassamp}

\end{document}